# GALExtin: An alternative online tool to determine the interstellar extinction in the Milky Way

Eduardo B. Amôres,[1]⋆ Ricardo M. Jesus,[2] André Moitinho,[3] Vladan Arsenijevic,[4] Ronaldo S. Levenhagen,[5] Douglas J. Marshall,[6,7] Leandro O. Kerber,[8] Roseli Künzel[5] and Rodrigo A. Moura[1]

[1] *Departamento de Física, Universidade Estadual de Feira de Santana (UEFS), Av. Transnordestina, S/N, CEP 44036-900 Feira de Santana, BA, Brazil*
[2] *PGCA-Universidade Estadual de Feira de Santana (UEFS)*
[3] *CENTRA, Faculdade de Ciências, Universidade de Lisboa, Ed. C8, Campo Grande, 1749-016 Lisboa, Portugal*
[4] *European Research Council Executive Agency, Brussels, Belgium*
[5] *Departamento de Física, Universidade Federal de São Paulo, Diadema, Brazil, Rua Prof. Artur Riedel, 275, 09972-270, Diadema, SP, Brazil*
[6] *Université de Toulouse, UPS-OMP, IRAP, F-31028 Toulouse cedex 4, France*
[7] *CNRS, IRAP, 9 Av. colonel Roche, BP 44346, F-31028 Toulouse cedex 4, France*
[8] *Departamento de Ciências Exatas e Tecnológicas, UESC, Rodovia Jorge Amado km 16, 45662-900, Brazil*

**ABSTRACT**
Estimates of interstellar extinction are essential in a broad range of astronomical research. In the last decades, several maps and models of the large scale interstellar extinction in the Galaxy have been published. However, these maps and models have been developed in different programming languages, with different user interfaces and input/output formats, which makes using and comparing results from these maps and models difficult. To address this issue, we have developed a tool called GALExtin (`http://www.galextin.org`) - that estimates interstellar extinction based on both 3D models/maps and 2D maps available. The user only needs to provide a list with coordinates (and distance) and to choose a model/map. GALExtin will then provide an output list with extinction estimates. It can be implemented in any other portal or model that requires interstellar extinction estimates. Here, a general overview of GALExtin is presented, along with its capabilities, validation, performance and some results.

## 1 INTRODUCTION

Estimation of interstellar extinction is key across many areas of Astronomy. From stellar studies where it affects the measurement of intrinsic fluxes, circumstellar reddening to the structure of our Galaxy where it affects the determination of distances and other properties, to extragalactic astronomy studies. In the latter case, it is necessary to estimate the contribution of the extinction of our Galaxy along all line of sight. All these investigations require knowledge of extinction towards given objects from their coordinates and distances (for 3D Models). Despite the difficulty in modelling extinction due to its clumpy distribution, many models and maps have been developed in the last three decades for describing the dust distribution in our Galaxy as 2D maps and 3D models. For detailed reviews of the studies regarding the modelling of Galactic extinction, we refer to the introductions of the works by Amôres & Lépine (2005); Marshall et al. (2006); Robin (2009); Nidever et al. (2012); Robin et al. (2015); Sale (2015); Green et al. (2015); Bovy et al. (2016).

Based on HI galaxy counts, Burstein & Heiles (1978, 1982) elaborated maps used to predict the integrated Galactic reddening covering a larger area in the sky. It was one of the most traditional and used maps until the end of the 1990s.

Schlegel et al. (1998) developed maps of the far-infrared emission from dust employing the COBE/DIRBE data (Kelsall et al. 1998). In Schlegel et al. (1998), reddening is determined both from the flux density at 100 $\mu$m and from a temperature correction map. Those maps have a massive usage in Astronomy with over 15,000 citations. Nevertheless, Arce & Goodman (1999), Chen et al. (1999) and Amôres & Lépine (2005), among other authors, point out the extinction values in those maps are overestimated towards some directions, mainly towards regions with $A_V > 0.5$ mag. Drimmel et al. (2003) formulated a model to describe the 3D extinction in the Galaxy considering the Galactic dust distribution of Drimmel & Spergel (2001) that uses of COBE/DIRBE data.

Amôres & Lépine (2005), hereafter AL, developed two models to describe the interstellar extinction in the Galaxy, considering the distribution for HI and H2. In their first model, the Galaxy is axisymmetric (ALA) and extinction grows linearly as a function of distance. In the second model (ALS), the extinction increases abruptly on paths passing through spiral arms. The representation of spiral arms is based on fitting $\ell$-v diagrams to the distribution of HI and HII regions and comparing the model density estimates as a function of Galactic longitude to those from HI surveys and IRAS fluxes at 100 $\mu$m as tracers of dust and CO, respectively. They also compared their models for a wide range of distances and directions to classical catalogues such as Neckel & Klare (1980), Savage et al. (1985) and Guarinos (1992). Amôres & Lépine (2005) estimated that, on average, Schlegel et al. (1998) overestimate the extinction by approximately 20%.

⋆ E-mail: ebamores@uefs.br (EBA)

Later, Amôres & Lépine (2007) revised their ALA model in order to achieve better estimates at intermediate and high Galactic latitudes. They calibrated their model with a sample of elliptical galaxies from Burstein (2003). They also compared their model predictions with open and globular clusters, as well as the extinction at the Galactic centre.

Dobashi et al. (2005) provided a map for $A_V$ covering Galactic longitudes (0° ≤ $\ell$ ≤ 360°) within ($|b|$ ≤ 40°) based on star counts. The maps are available in both high (6') and low (18') resolution. Afterwords, using 2MASS star counts and the BGM, Dobashi et al. (2013) render maps with two resolutions, 1' and 15', respectively.

Marshall et al. (2006) used the Besançon Galaxy Model (Robin et al. 2003, 2016; Amôres et al. 2017), hereafter BGM, and 2MASS data (Skrutskie et al. 2006) to build a model aiming at describing interstellar extinction in 3D. The map coverage is $|\ell|$ ≤ 100° for $|b|$ < 10° for several lines of sight.

Chen et al. (2013) assumed a similar method and combined the VVV[1] (Saito et al. 2012), 2MASS and GLIMPSE data (Churchwell et al. 2009) to derive the extinction towards the Galactic bulge for planar regions, i.e., $|\ell|$ ≤ 10° for $|b|$ ≤ 2°. In a following work, the authors (Schultheis et al. 2014) extended the latitude coverage to -10° < b < 5°.

Rowles & Froebrich (2009) developed All-Sky extinction maps $|b|$ ≤ 30° based on the median near-infrared colour excess technique with 2MASS data.

Gonzalez et al. (2011) elaborated a method for obtaining reddening maps by using the mean $J - K_s$ colour of red clump (RC) stars based on VVV data. Later, Gonzalez et al. (2012) applied this method to the entire bulge observed by VVV deriving maps with a resolution of 2' and 6'.

Schlafly et al. (2010) checked the reddening law and the accuracy of the Schlegel et al. (1998) maps employing SDSS photometry (Abazajian et al. 2009). They reported that Schlegel et al. (1998) overestimated E(B-V) by 14%. Later, Schlafly & Finkbeiner (2011) validated this result and calibrated the Schlegel et al. (1998) maps applying the Fitzpatrick & Massa (2005) extinction law and SDSS data. They also implemented a handy table (Table 6 in the publication) to achieve the conversion from $E(B - V)_{SFD}$ to $E(B - V)$ for roughly one hundred passbands and four $R_V$ values.

Using the Xuyi Schmidt Telescope Photometric Survey (Zhang et al. 2014) of the Galactic Anticentre, Chen et al. (2014) elaborated a 3D extinction map in the $r$ band for the Galactic anti-centre with coverage for Galactic longitude 140° < $\ell$ < 240° and latitude -60° < b < 40°.

Schlafly et al. (2014), based on a sample of 500 million of stars (PS1) for the northern sky, derived a map of dust reddening up to 4.5 kpc for two resolutions, 7 and 14 arcmins.

Sale et al. (2014) handled IPHAS (Drew et al. 2005) data with a hierarchical Bayesian model (Sale 2012) to derive interstellar extinction for the entire northern Galactic plane 30° ≤ $\ell$ ≤ 230° ($|b|$ < 5°).

The observations performed by the Planck satellite (Planck Collaboration et al. 2014a) have been important for modelling and obtaining All-Sky parameters of thermal dust emission (Planck Collaboration et al. 2014b). Furthermore, Planck Collaboration et al. (2014b) provided the values of several properties of dust emission, namely, the dust temperature, the dust optical depth at 353 GHz ($\tau_{353}$), the dust spectral index, the dust radiance and the E(B-V) estimates for extra-galactic studies, among others. The Planck´s Dust Model Page (see Table 1) contains individual files for each of those five parameters.

Green et al. (2015) presented a 3D interstellar extinction map based on Pan-STARSS 1, hereafter PS1 (Kaiser et al. 2010), and 2MASS data for three-quarters of the sky. Using a similar method, Green et al. (2018) produced a new 3D map of interstellar extinction based on stellar photometry from PS1 and 2MASS. The map covers approximately three-quarters of the sky (declinations with magnitudes for $\delta \geq -30°$). They also elaborated on another 3D map (Green et al. 2019) based on the Gaia-DR2 (Gaia Collaboration et al. 2018) that covers the sky north of a declination of -30°. The results can be obtained through an interface (as shown in our Table 1) written in Python. They also provide a catalogue containing distances, reddenings and types of 799 million stars. In the work by Hanson et al. (2016), the authors use both PS1 and Spitzer/GLIMPSE (Churchwell et al. 2009) data to develop a 3D map with angular resolution of 7 arcmin towards 0° < $\ell$ < 250° for $|b|$ ≤ 4.5°.

Capitanio et al. (2017) developed 3D extinction maps based on a regularized Bayesian inversion of individual colour excess data. In a previous work (Lallement et al. 2014), the authors developed a 3D model based on the inversion method employing a sample of 23,000 stars within 2.5 kpc. Later, Lallement et al. (2019) rendered 3D maps based on both 2MASS and Gaia-DR2 data applying Bayesian inversion.

Chen et al. (2019) elaborated a 3D map based on 2MASS, Gaia-DR2, and spectroscopic survey data using a machine-learning algorithm (Random Forest regression).

Soto et al. (2019) elaborated a map for a colour-excess extinction using VVV and GLIMPSE data towards the fourth quadrant, 295° ≤ $\ell$ ≤ 350° at Galactic latitudes ($|b|$ < 1°) in some cases going up to $|b|$ < 2° covering 148 square degrees. The authors used the Rayleigh-Jeans Colour Excess (RJCE) technique.

Babusiaux et al. (2020) developed a Bayesian deconvolution method to compute distance and interstellar extinction by using catalogues of reference. In another work, they applied (Hottier et al. 2020) the technique called FEDReD to elaborate a map of interstellar extinction based on 2MASS and Gaia-DR2, that covers the entire Galactic disc ($|b| \leq 0.24^o$) reaching distances up to 5 kpc in the direction of the Galactic centre.

Guo et al. (2020) elaborated a 3D interstellar extinction map for the southern sky covering 14,000 square degrees. They used the SkyMapper Southern Survey (Wolf et al. 2018) with Gaia-DR2 and 2MASS and WISE data.

Regarding the estimation of interstellar extinction by users, there are a number of tools for computing extinction either as programs to run on the user's computer or as on-line services. Table 1 gives a list of these services as well as their links.

Those services are in general based on maps or models de-veloped in different programming languages (that use specific file annotations, some of them large ASCII tables), with different user interfaces and input/output formats. The variety of interfaces and formats create difficulties on the user side, in practice limiting the extinction maps and models used in scientific analysis. This in turn limits the ability of comparing the effects of using different maps and models. It can also lead to the use of maps or models that may not be the most appropriate option for study being made.

---

[1] VISTA Variables in the Via Lactea

Here we present a web-based tool, GALExtin. The service has been created to overcome the problem above, by integrating several maps and models in a unique tool using one common simple user interface and output format.

The remainder of this paper is organised as follows. Section 2 presents an overview of available interstellar extinction tools. GALExtin with the models/maps it provides is presented in Section 3. A comparison among extinction estimate obtained with some models/maps is presented in Section 4. Section 5 addresses the conclusions of this study and gives some final remarks.

## 2 A GENERAL OVERVIEW OF THE AVAILABLE INTERSTELLAR EXTINCTION TOOLS

The usage of either an interstellar extinction map or model depends directly not only on its capabilities but also its availability and feasibility. Table 1 shows several available online services and repositories that provide the interstellar extinction of several models and maps. Most of them are based only on the estimates provided by their own authors. In the following section, we will briefly describe each of them and provide the respective links to the interested reader.

By the beginning of the 1980s, the maps of Burstein & Heiles (1978, 1982) were distributed as ASCII files (the FORTRAN program outputs), or as contours maps in the printed AJ papers. Further authoritative catalogues are those provided by Neckel & Klare (1980) and Fitzgerald (1968) which yield E(B-V) towards a large number of stars.

In the NED/IPAC webpage[2], there is also a service called "Coordinate Transformation & Galactic Extinction Calculator". This service provides not only transformation between different coordinates systems but also interstellar extinction as provided by Burstein & Heiles (1978, 1982) and Schlegel et al. (1998), as well as its value using the calibration values provided by Schlafly & Finkbeiner (2011).

Arenou et al. (1992) presented a 3D model that uses a sample of 215,000 stars contained in the INCA Database (Turon et al. 1991). On the extinction model page, users can choose from three methods: i) extinction from $V$ mag and $B - V$, ii) extinction of $V$ mag and iii) extinction as function of $\ell$, $b$ and distance.

Since their publication, the models of Amôres & Lépine (2005) were provided in the IDL language or in FORTRAN[3] as The Model ALA. In a previous version of the current project, they were also available for a private list of collaborators with a restricted list of maps and models (Amôres et al. 2012).

The maps of Dobashi et al. (2005, 2013) are available through an interactive interface (for the link see Table 1), with an option to download a FITS file for a given region of the sky.

Rowles & Froebrich (2009) launched an online interface where they also provide maps for three different resolutions as well as E($B - V$) for Schlegel et al. (1998) and Dobashi et al. (2005). Furthermore, it is also feasible to retrieve extinctions for smaller-scaled maps as those produced by Lombardi et al. (2006, 2008); Cambrésy (1999).

It is also possible to obtain the extinction from the maps of Gonzalez et al. (2011, 2012) for a given pair of coordinates through an interactive interface (BEAM Calculator, see also Table 1) in which the users can choose from two different extinction laws, either the one from Cardelli et al. (1989) or from Nishiyama et al. (2009).

An online tool (see also Table 1) retrieves the data from Schlafly et al. (2014) which allows the user to obtain the E(B-V) value for a given pair of $\ell$,$b$. The user can also download the map, including instructions on how to read it using either IDL or Python code.

The data from Planck Collaboration et al. (2014b) can also be obtained in a single file called[4] available in the Planck Legacy Archive[5].

Morales-Durán et al. (2015) developed a tool in the context of Spanish Virtual Observatory that provides $A_V$, $R_V$ and distance for a sample of approximately 200,000 stars. Their service provides a list of stars for a given pair of coordinates and search radius supplied by the user.

Bovy et al. (2016) in the context of upcoming near-infrared spectroscopic surveys and Gaia (Gaia Collaboration et al. 2016) developed a likelihood-based phase-space inference framework for the MW. They also combined extinction models/maps (Marshall et al. 2006; Drimmel et al. 2003; Sale et al. 2014; Green et al. 2015, 2018) in a tool developed in Python and available for download (see Table 1).

The maps of Hanson et al. (2016) cover an area over 7,000 square degrees towards the Galactic plane with a resolution of 6 arcmins and are displayed as a large table (see Table 1).

Green (2018) elaborated a unified Python interface, called *dustmaps* (Table 1), to retrieve interstellar extinction from several models and maps. This interface also includes the maps by Green et al. (2015, 2018, 2019) designated as Bayestar15, Bayestar17 and Bayestar19, respectively.

Santiago et al. (2016) derived a method to compute spectrophotometric distances applying a Bayesian approach and SEGUE and APOGEE data (Allende Prieto et al. 2008). Later on, they developed (Queiroz et al. 2018) and applied their method (called StarHorse) to compute distances and extinction using APOGEE, RAVE (Steinmetz et al. 2006), Gaia-ESO Survey (GES, Gilmore; 2012). (Anders et al. 2019) using Gaia-DR2 combined with PS1, 2MASS, and AllWISE, computed not only stellar parameters but also distances and extinction for 265 million stars brighter than $G$=18.

Lenz et al. (2017) presented a map of interstellar reddening based on HI column densities for 39% of the sky with good accuracy in the low density regime.

The map provided by Lallement et al. (2019) covers distances up to 3 kpc and is available for use either through their tool (Table 1) or via download at CDS[6]. In addition, the 3D extinction maps provided by (Lallement et al. 2014) and Capitanio et al. (2017) are also available from an on line interface (Table 1).

---

[2] https://ned.ipac.caltech.edu/extinction_calculator
[3] http://galextin.org/amores_lepine.php
[4] HFI_CompMap_ThermalDustModel_2048_R1.20.fits
[5] http://pla.esac.esa.int/pla/#home
[6] https://cdsarc.unistra.fr/viz-bin/ReadMe/J/A+A/625/A135?format=html&tex=true

Soto et al. (2019) set a service that provides extinction for one pair of coordinates (see Table 1). It is also possible to download maps and to use an IDL script to retrieve extinction values at specified coordinates. The map of Babusiaux et al. (2020) is available at the following link[7]. The map of Guo et al. (2020) is availableat this link[8], with a given example of python code to access the interstellar extinction from a 3D position.

## 3 GALEXTIN

GALExtin can either deliver the interstellar extinction or reddening towards any given position in our Galaxy providing the coordinates and distance, or an integrated extinction along a line of sight. In the latter case, more useful for extragalactic studies, only coordinates need to be provided, since they are 2D estimates (Burstein & Heiles 1978, 1982; Schlegel et al. 1998; Schlafly et al. 2014; HI4PI Collaboration et al. 2016; Lenz et al. 2017).

The user must provide either Galactic or Equatorial coordinates ($\ell$,$b$) of a single object (direction) or a list of coordinates (including distance) in an ASCII file separated by spaces. Presently, only lists with up to 1,000 directions and for one model/map per run are accepted.

The GALExtin front-end is designed as a web application, with a network background composed of HTML tags, PHP scripts and access to a MySQL database. We wrote the majority of the programs in the IDL language, with minor subroutines in Python, FORTRAN and C.

GALExtin works with two layers. The first one is a client that provides a web form to be filled out by the users, e.g. data such as coordinates and distance for the 3D extinction. Alternatively, the user can insert a list of coordinates (Galactic or Equatorial) with distances. It is also necessary to select the coordinate system and the desired model/map.

The PHP program, embedded in the HTML code, accesses an SQL table to attribute a number to the process, unique identification for each run of GALExtin, which is also used to assign and manipulate input and output files names.

Once PHP receives a process number and the information passed through HTML, it calls an IDL program that manages the extinction computation. It verifies the number of input lines. Its primary task consists of calling the routines that compute the interstellar extinction for each chosen model/map.

In the final step, the routine returns the extinction estimates to the main program. The output is displayed in the web form if a single direction is given, or written to a file if a file with coordinates was given. For either Galactic or Equatorial coordinates, the values must be given in decimal degrees.

### 3.1 Models available in GALExtin

The two models for interstellar extinction determination in the Galaxy presented by Amôres & Lépine (2005) are available: the Axysimetric (Model A) and the Spiral Model (Model S) as well as one (Model A2) described in Amôres & Lépine (2007) that improves a little more the Axysimetric Model outside the Galactic plane. The three models are written in IDL.

---

[7] http://cdsarc.u-strasbg.fr/viz-bin/cat/J/A+A/641/A79
[8] https://nadc.china-vo.org/article/20200722160959?id=101032.

For Burstein & Heiles (1978, 1982), we used the bhrdfort.pro routine that was written by D. Schlegel based on Michael Strauss' FORTRAN data files that return as output 4×E($B-V$). When using those values for E($B-V$), attention should be taken to the zero-points (Burstein 2003). As pointed out in the documentation of the routine, no data means -14, -99 or -396 and valid data has the output > -0.22.

The most used extinction estimates in the literature are those of Schlegel et al. (1998), which are also available in GALExtin. We have used the function dustgetval.pro with interpolation. GALExtin provides not only the E($B-V$)$_{SFD}$ but also the A$_V$ for Landolt $V$ (R$_V$ = 3.1) by using Fitzpatrick & Massa (2005) reddening law as presented by Schlafly et al. (2010) (Table 6). In the case of other bandpasses, it is only necessary to multiply E($B-V$)$_{SFD}$ by the value provided in this table. For both Burstein & Heiles (1978, 1982) and Schlegel et al. (1998), we have use the fits, routines and data files provided at[9].

The 3D Model of Drimmel et al. (2003) provides the extinction in the $V$ band. We got the FORTRAN version provided on the author's webpage[10].

The extinctions provided by Marshall et al. (2006)[11], Chen et al. (2013)[12], Chen et al. (2014); Sale et al. (2014) are all given in tables with coordinates information followed by ranges of distance, extinction and the errors. The links to obtain the extinction for Chen et al. (2014) and Sale et al. (2014) are provided in Table 1.

To better manipulate those auxiliary files, we split each one of them into two fits files, the first one with the coordinates and the second with a single structure containing extinction, distance and the error for a given distance, as well as the number of the field that is linked with the coordinates file.

For instance, the Marshall et al. (2006) original file has a size of 52 MB. It is split in two files, one with the coordinates and the other with interstellar extinction and distance estimates. The sizes of the split files are approximately 0.5 and 8.3 MB. The elapsed time to read those fits files is almost 1% of the time to read the large Marshall et al. (2006) table. One important characteristic of GALExtin is that for each process, the large files as well as the main data in the models, are loaded only once in memory.

From the coordinates ($\ell$,$b$) for a given extinction, GALExtin gets the nearest position allowing a maximum distance of 0.5 degree. In the output, not only the input but also the obtained ($\ell$,$b$) and distance are provided, which informs the user how far from the original direction the output is.

As the intervals between the points in the extinction data from Marshall et al. (2006); Chen et al. (2013) could be large (for instance, the distance between the last two points is 1.0 kpc) we performed a linear interpolation between the points to provide a finer estimate of the interstellar extinction and the error, rather than use the estimate given by the authors. The interval for the interpolation in the distance is equal to 10 pc and 50 pc for Marshall et al. (2006) and Chen et al. (2013), respectively. Regarding the map of Chen et al. (2013), the colour excess for both E($J-Ks$) and E($H-Ks$) are provided.

In the case of 3D data obtained from Chen et al. (2014) we do not interpolate the data since the resolution of their map is 0.15 kpc.

---

[9] https://doi.org/10.7910/DVN/EWCNL5
[10] ftp://ftp.oato.inaf.it/astrometria/extinction
[11] https://cdsarc.unistra.fr/viz-bin/ReadMe/J/A+A/453/635?format=html&tex=true
[12] https://cdsarc.unistra.fr/viz-bin/ReadMe/J/A+A/550/A42?format=html&tex=true

In this case, for a given pair of coordinates, we consider the point nearest to the input direction. The estimates of Chen et al. (2014) reach distances up to 4.35 kpc; the authors present extinction in $A_r$ that can be converted to $A_V$ using the linear relation $A_V = 1.172 \times A_r$ (Yuan et al. 2013).

For Sale et al. (2014) despite the resolution in the distance being equal to 0.10 kpc, we still interpolate it with intervals equal to 0.05 kpc. We applied the interpolation to compare it with the data obtained (see Section 3.2) from CDS-Vizier available for distances equal to 1, 3 and 10 kpc. As the original map's first extinction value is for a distance equal to 0.05 kpc (with steps of 0.10 kpc), we only obtain those distance values by interpolating.

The Planck Dust Model for All-Sky is available in the GALExtin for E(B-V) for extragalactic studies. It is essential to mention that both values are integrated, i.e., along with all the MW line of sight. We also provide the HI column density ($N_{HI}$ in cm$^{-2}$) obtained from the observations of HI4PI Collaboration et al. (2016) available in FITS format at the CDS.[13]

The maps presented in Sale et al. (2014) cover the Galactic plane towards the Northern sky with distances up to 15.0 kpc. They provide the monochromatic extinction ($A_0$) at 5495 Å. As pointed out in Sale et al. (2014), $A_0 = 1.003 \times A_V$.

We also provide the interstellar extinction given by *dustmaps* from Green et al. (2015, 2018, 2019) using the mode equal to best for Bayestar15, Bayestar17 and Bayestar19. Programs and data from *dustmaps* are only used by GALExtin for this purpose. The output values are given in specific units that differ among them Green et al. (2018, 2019). To convert them to either E(B-V) or an extinction in any passband, see the those papers and the links[14] and [15].

The Colour-Excess map by VVV (Soto et al. 2019) is also available within GALExtin, with $A_K$ values and their standard deviations. The maps of interstellar reddening by Lenz et al. (2017) are available in GALExtin with values provided in $E(B-V)$.

Chen et al. (2019) provided, in a single file, the colour excesses as a function of the Galactic longitude, latitude and distances, ranging from 0.2 to 6.0 kpc with a step of 0.2 kpc. Within GALExtin, we have split the extinction data by Chen et al. (2019) into two files. One file has the Galactic coordinates and maximum distance for each pair of coordinates. The second file has a key to assign the distance and the extinction in $A_G$ with its error estimate. We used the expressions provided by the authors to compute $A_G = E(G\text{-}K_S) + 1.987 \times E(H\text{-}K_S)$. The error was estimated applying $A_G = 1.987 \times E(H\text{-}K_S)_{error}$. GALExtin provides the extinction of the point nearest to the input distance.

Figure 1 shows the values obtained using GALExtin and those by other authors for some line of sights. As can be seen, there is full agreement between the GALExtin values and the results provided by the other authors.

### 3.2 Validation

The maps and models currently available in GALExtin were validated in the sense that results obtained with GALExtin are fully reproduced, which can be seen by comparing with the original data provided by the authors, with *dustmaps*, and other ways, as described below. The fact that in the context of GALExtin only the code available for the maps of Bayestar15, Bayestar17 and Bayestar19 is used from *dustmaps*, reinforces the robustness of our results. On the GALExtin's page, we also provide detailed information for all comparisons and files used on them.

It is essential to verify the reproducibility and consistency of the results. To do so, we performed some comparisons based on the model/map coverage. For instance, in the maps (Burstein & Heiles 1978, 1982), we have compared the results obtained using GALExtin and the obtained values by *dustmaps* towards a sample of 378 elliptical galaxies as published by Burstein (2003).

For the map of Schlegel et al. (1998), we have used a grid of points based on Hierarchical Triangular Mesh (HTM) level 6 (Kunszt et al. 2001) to get the pairs of coordinates distributed over All-Sky, 32,768 points in total. In this sense, we have compared the results obtained by GALExtin and also the values obtained using *dustmaps*.

We used the same grid with added distance, ranging randomly, from 0 to 30 kpc for the models of Amôres & Lépine (2005, 2007), except for the Model S in which we reduced the tests for the inner Galaxy ($|\ell| \leq 100.0^o$) for $|b| \leq 10.0^o$ totalizing 12,598 points. For those models, we have used the same versions provided on their webpage and compared the results obtained by using the GALExtin with the values obtained on a local machine.

Concerning the Model of Drimmel et al. (2003) there is a file in its webpage (see above) with interstellar extinction for eight directions, we have compared the values obtained using GALExtin not only for those directions but also for a grid with 2,000 pairs of coordinates (see below) using GALExtin and the same author's code on a local machine.

For the All-Sky Planck Dust Model we have compared the values obtained with GALExtin for E(B-V), but we also built from the HTM level 6 another grid of points containing 2,000 pairs of coordinates distributed at random for All-Sky, comparing the results from GALExtin with the values obtained using dustmaps interface.

For Marshall et al. (2006); Chen et al. (2013); Sale et al. (2014) we have downloaded the data from CDS-Vizier interface application considering a regular interval of $\ell, b$ and distance. An important aspect resides in the fact that the original values for that extinction estimates were also obtained from CDS (in a single file), but in this case, we downloaded the entire full file and also used another program to read the files for avoiding any bias. For Chen et al. (2013) we performed the validation with both $E(J - K)$ and $E(H - K)$.

To validate those three maps, we have compiled a catalogue of 7,190; 711 and 1,050 pairs of coordinates and distance. For Chen et al. (2014), we got points randomly from the data file provided by authors using a procedure different than the one used by GALExtin to fetch its data.

To check the results provided by Bayestar (Green et al. 2015, 2018, 2019), we ran GALExtin for some directions as described in the documentation of *dustmaps* as well as comparing the values using GALExtin with those obtained running bayestar local for the same file with 2,000 pairs of coordinates.

To verify the HI4PI collaboration data, we have also used the CDS-Vizier interface application considering fetching HI column density for Galactic longitudes ranging from 0.0° to 355.0° (steps of 20.0 degrees) and Galactic latitudes from -85 to +85 degrees (steps of 10.0 degrees), totalizing 660 pairs of coordinates.

We validated the map by Lenz et al. (2017) in two different ways. In a first approach, we considered four pairs of Galactic longitude and latitude and $E(B-V)$ as provided by the authors in their web-site (see Table 1). Later, we considered a list of 2,000 pairs of coordinates running in the GALExtin as well as with *dustmaps*. For the record, there are 754 pair of coordinates with a valid $E(B-V)$ estimate towards the list above.

---

[13] ftp://cdsarc.u-strasbg.fr/pub/cats/J/A+A/594/A116
[14] http://argonaut.skymaps.info/usage#units
[15] https://dustmaps.readthedocs.io/en/latest/modules.html

Using the Chen et al. (2019) original reddening file, we produced a list of 1,000 directions with Galactic coordinates and then compared with the GALExtin results. In all comparisons, we find the relative difference between the results from GALExtin and those provided in the comparison files to be equal to zero (see Table 2 for the r.m.s.). It is worth noticing a minimal difference of $2.951\times 10^{-3}$ in the r.m.s. in comparison with the $A_0$ value provided by Sale et al. (2014) between CDS and our values for the three ranges of distances returned by CDS at 1, 3 and 10 kpc. This small difference comes from the fact that we interpolated the values of extinction along the line of sight. We also verified the values obtained by GALExtin for Sale et al. (2014) for a list with 5,508 points, containing 612 pairs of $(\ell,b)$ for nine distances, obtained aleatory directly from IPHAS: 3D extinction map[16]. We notice that the r.m.s. ($5.042\times 10^{-5}$) in this comparison is lower than the obtained one in the previous comparison (see Table 2).

To validate the map of Soto et al. (2019), we elaborated a grid with 555 points covering its whole region, at every 0.5 degrees for both Galactic longitude and latitude. The r.m.s. is equal to 0.0003.

For the record, Table 2 also shows the elapsed time for each model/map taking into account a given number of lines of sight. As can be seen, even for a large number of lines of sight, the computation time is low. We performed these simulations running GALExtin only on our server.

## 4 RESULTS

In addition to the sanity check reported in the previous section, we also expand our comparison of the results obtained with GALExtin and other authors in the same way performed by Bovy et al. (2016) in their Fig. 5. The authors compare, in this figure, the values of $A_H$ provided by several authors as a function of the distance.

In order to compare the estimates of GALExtin, we have elaborated a similar figure also adding the estimations of Amôres & Lépine (2005) models. Figure 2 shows this comparison. For most of directions we have considered, the Axis and Spiral of Amôres & Lépine (2005), except for directions in which the contribution of spiral arms is is either not expected or is negligible.

Differently from Bovy et al. (2016), who considered an average extinction within a radius centred on the coordinates, we only consider the extinction towards a given direction (top in the panels (Figure 2). Similar to Bovy et al. (2016), we also got the stars nearest to the centre of the coordinates; in our case, stars from the APOGEE-RC Catalogue[17] within a box of size 1° x 1°. To convert values from $A_J$ to $A_H$ and $A_K$, we have used the values provided by Rieke & Lebofsky (1985), i.e., 0.175 and 0.112, respectively.

In general, there is a good agreement between the models in all directions, and at least for some distance ranges. The differences can be attributed to the characteristics of each model, their coverage, resolution, and traces used to elaborate them. We can see an interesting feature in the Amôres & Lépine (2005) spiral model, which is the fact that extinction grows by discrete steps passing through spiral arms. Furthermore, model is in good agreement with the extinction of APOGEE-RC Catalogue. We yield Amôres & Lépine (2005) - Model A, as the first option to retrieve using the fine-tuning, in the direction of $(\ell,b) = (127.5°, 0.0°)$, however GALExtin provides extinction values also considering no fine-tuning option.

## 5 CONCLUSION

We developed an online tool that supports interstellar extinction estimates for twenty models/maps available in the literature. The user only needs to supply a pair of coordinates and possibly the distance, or alternatively, a file with multiple sky coordinates. We validated our tool comparing the results of GALExtin with the ones obtained with the original software of the models, *dustmaps*, as well as with the values obtained using the Vizier–CDS in the case of some maps. We plan to include new models/maps in GALExtin regularly. In the future, we also plan to have VO accessibility and interfaces and use VOTable as default data format.

By providing a uniform web-based interface and output format, GALExtin allows users to explore, compare and otherwise use several extinction models and maps, without having to install many different programs or even download large tables. We wish to emphasize that GALExtin builds on the the work done by other authors who have developed the underlying tools and maps, and that there are several other excellent extinction maps and models not currently offered in GALExtin. To clearly acknowledge these works, brief descriptions, as well as links to their web pages are presented in Table 1.

## ACKNOWLEDGEMENTS

We thank the anonymous referee for his/her beneficial suggestions that helped us improve the manuscript and the GALExtin service. Eduardo Amôres thanks Dr Annie Robin, Dr Beatriz Barbuy, Dr Claudia Vilegas, Dr Jacques Lépine, Dr Francesca Figueras, Dr Laerte Sodré Jr., Dr Joachim Köppen and Dr Kevin Alton for the the encouragement and also Dr Angelo Duarte to let his machine also partially available in the initial phase of this project. Disclaimer: The views expressed in this paper are the authors' (VA). They do not necessarily reflect the views or official positions of the European Commission, the ERC Executive Agency or the ERC Scientific Council. AM acknowledges support from the Portuguese Fundação para a Ciência e a Tecnologia (FCT) through the Strategic Programme UID/FIS/00099/2019 for CENTRA. RAM thanks to PIBIC/CNPq for the scholarship (148148/2019-0 and 159001/2020-0). This research has made use of the VizieR catalogue access tool, CDS, Strasbourg, France. This research also made use of Astropy, a community-developed core Python package for Astronomy(Astropy Collaboration et al. 2013, 2018).

## DATA AVAILABILITY

The data underlying this article were accessed from http://www.galextin.org. The derived data generated in this research will be shared on reasonable request to the corresponding author.

---

[16] http://www.iphas.org/data/extinction/Amap.tar.gz
[17] http://www.sdss.org/dr14/data_access/value-added-catalogs/?vac_id=apogee-red-clump-rc-catalog

8   *E. B. Amôres et al.*

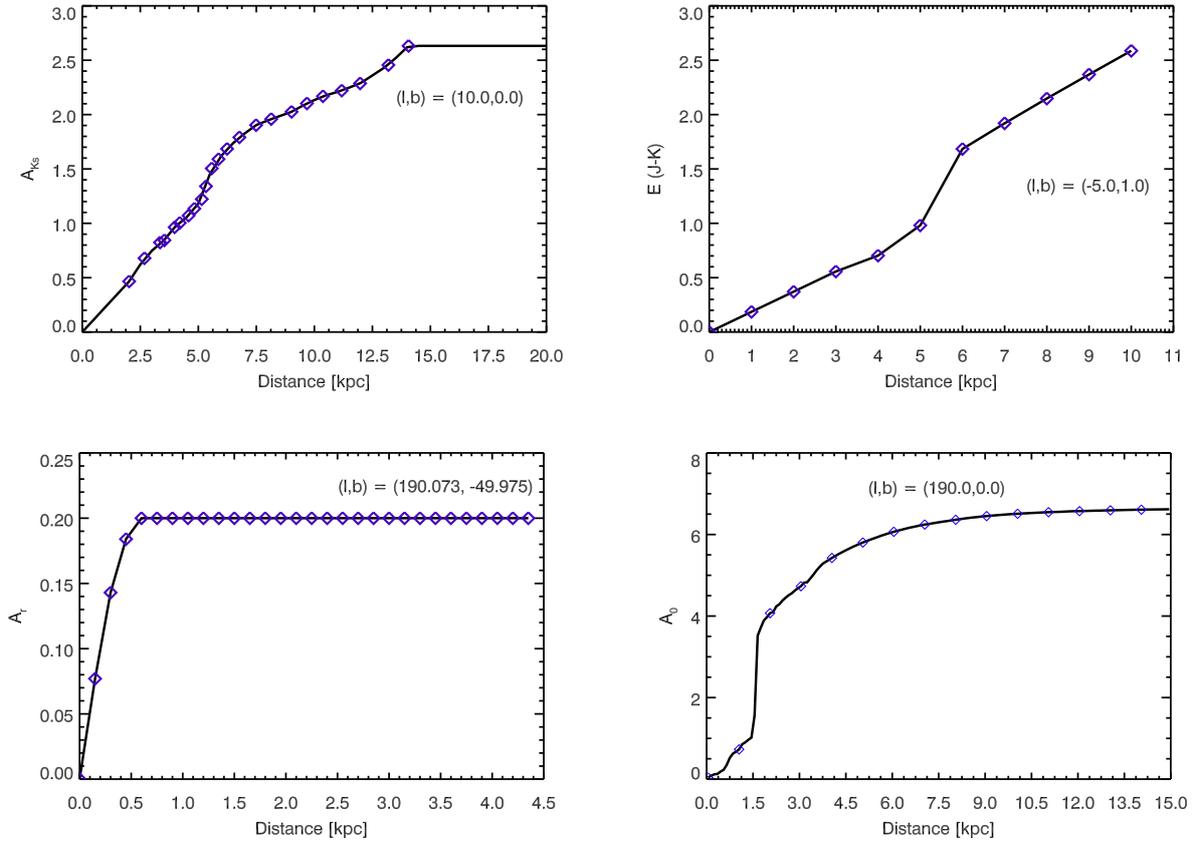

**Figure 1.** Extinction versus distance as provided for some authors (points): upper right panel (Marshall et al. 2006); upper left panel (Chen et al. 2013); lower right panel (Chen et al. 2014); lower left panel (Sale et al. 2014) and the obtained ones using GALExtin (lines). The pairs of coordinates are indicated at the top of each panel.

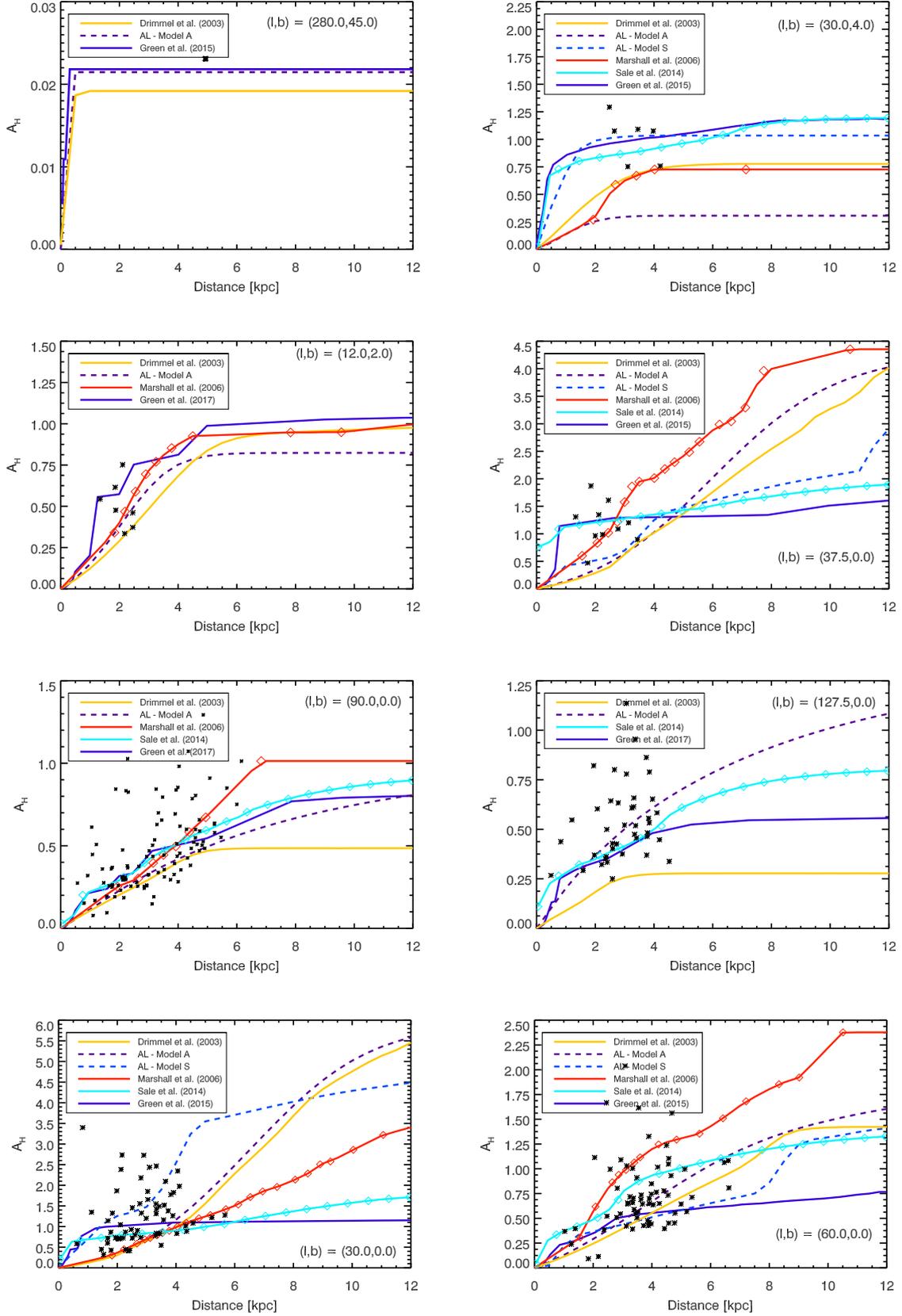

**Figure 2.** Extinction versus distance for some models as showed in the legends. The asterisks are stars of APOGEE-RC Catalogue located within four square degrees of the center of the coordinates presented at each panel.

**Table 1.** List of available sites and services that estimate the interstellar extinction in our Galaxy.

| Service/repository | link |
| --- | --- |
| Interstellar extinction using a 3D Galactic Model (Arenou et al. 1992) | `http://wwwhip.obspm.fr/cgi-bin/afm` |
| Galactic Extinction Calculator (Schlegel et al. 1998; Schlafly & Finkbeiner 2011) and coordinate transformation | `https://ned.ipac.caltech.edu/forms/calculator.html` |
| 2D maps based on DSS and 2MASS data (Dobashi et al. 2005; Dobashi 2011; Dobashi et al. 2013) | `http://darkclouds.u-gakugei.ac.jp/` |
| Models for interstellar extinction in the Galaxy (Amôres & Lépine 2005) | `http://galextin.org/amores_lepine.php` |
| All sky maps (Rowles & Froebrich 2009; Dobashi et al. 2005; Schlegel et al. 1998) and smaller scale maps (Lombardi et al. 2006, 2008; Cambrésy 1999) | `http://astro.kent.ac.uk/~df/query_input.html` |
| BEAM extinction calculator (Gonzalez et al. 2011, 2012) | `http://mill.astro.puc.cl/BEAM/calculator.php` |
| RJCE Two-Dimensional Extinction Maps | `https://www.noao.edu/noao/staff/dnidever/rjce/extmaps/index.html` |
| Planck´s Dust Model Page | `http://hyperstars.fr/mamd/planck_dust_model.html` |
| Repository of A 3D extinction map of the Northern Galactic Plane based on IPHAS photometry (Sale 2012; Sale et al. 2014) | `http://www.iphas.org/extinction/` |
| Repository of Photometric Extinctions and Distances (Chen et al. 2014) | `http://162.105.156.249/site/Photometric-Extinctions-and-Distances/` |
| A Map of Dust Reddening to 4.5 kpc from Pan-STARRS1 (Schlafly et al. 2014) | `http://faun.rc.fas.harvard.edu/eschlafly/2dmap/querymap.php` |
| Spanish VO - Extinction Map (Morales-Durán et al. 2015) | `http://svo2.cab.inta-csic.es/theory/exmap/` |
| A Three-dimensional Map of Milky Way Dust (Green et al. 2015, 2018, 2019) | `http://argonaut.skymaps.info` or `https://dataverse.harvard.edu/dataset.xhtml?persistentId=doi:10.7910/DVN/2EJ9TX` or |
| Dustmaps: A Python interface for maps of interstellar dust (Green 2018) | `https://dustmaps.readthedocs.io/en/stable/index.html` or `https://github.com/gregreen/dustmaps` |
| On Galactic density modeling in the presence of dust extinction (Bovy et al. 2016) | `https://github.com/jobovy/mwdust` |
| A new, large-scale map of interstellar reddening derived from HI emission (Lenz et al. 2017) | `https://github.com/daniellenz/ebv_tools` |
| Structuring by Inversion the Local ISM (Lallement et al. 2014; Capitanio et al. 2017) | `http://stilism.obspm.fr/` |
| Gaia/2MASS: 3D Dust Extinction Map (Lallement et al. 2019) | `https://astro.acri-st.fr/gaia_dev/about/` |
| Dust in three dimensions in the Galactic plane (Hanson et al. 2016) | `http://www.rhanson.de/gpdust/` |
| StarHorse: a Bayesian tool for determining stellar masses, ages, distances, and extinctions for field stars (Queiroz et al. 2018) | `http://www.linea.gov.br/020-data-center/acesso-a-dados-3/acesso-a-dados-2/#spectrophotometric_distances_and_extinction` |
| Three-dimensional interstellar dust reddening maps of the Galactic plane (Chen et al. 2019) | `http://paperdata.china-vo.org/diskec/cestar/table1.zip` |
| Photo-astrometric distances, extinctions, and astrophysical parameters for Gaia DR2 stars brighter than $G$ = 18 (Anders et al. 2019) | `https://gaia.aip.de/` |
| A Color-Excess Extinction map of the Southern Galactic disk from the VVV and GLIMPSE Surveys Soto et al. (2019) | `http://astro.userena.cl/ExtMapVVV/` |

This paper has been typeset from a TeX/LaTeX file prepared by the author.

**Table 2.** Models/maps available in GALExtin with the computation time in our server for a given line of sights.

| Model/map | Time | number of los | reddening/extinction | r.m.s. |
|---|---|---|---|---|
| Burstein & Heiles (1978, 1982) | 2 s | 378 | $E(B-V)$ | $2.770 \times 10^{-3}$ |
| Schlegel et al. (1998) | 1 m | 32,768 | $E(B-V)_{SFD}$ and $V$-Landolt (Schlafly et al. 2010) | $1.370 \times 10^{-4}$ |
| Drimmel et al. (2003) | 1 m | 2,000 | $A_V$ | $1.100 \times 10^{-7}$ |
| Amôres & Lépine (2005) – AL1 | 22 s | 32,768 | $A_V$ | 0.000000 |
| Amôres & Lépine (2005) – ALS | 7 m | 12,598 | $A_V$ | $1.992 \times 10^{-5}$ |
| Amôres & Lépine (2005) – ALNF | 22 s | 32,768 | $A_V$ | $7.813 \times 10^{-6}$ |
| Amôres & Lépine (2007) – AL2 | 41 s | 32,768 | $A_V$ | $1.104 \times 10^{-5}$ |
| Marshall et al. (2006) | 4 m | 7,190 | $A_K$ | 0.000000 |
| Chen et al. (2013) | 5 s | 711 | $E(J-K)$ and $E(H-K)$ | 0.000000 for both |
| Chen et al. (2014) | 70 s | 711 | $A_V$ | 0.000000 |
| Sale et al. (2014) | 30 s | 1,050 | $A_0$ at 5495Å | $2.951 \times 10^{-3}$ and $5.042 \times 10^{-5}$ |
| Schlafly et al. (2014) | 6 s | 100 | $E(B-V)$ | 0.000000 |
| Planck Collaboration et al. (2014b) | 10 s | 2,000 | $E(B-V)$ | 0.000000 |
| HI4PI Collaboration et al. (2016) | 4 m | 660 | $N_{HI}$ in cm$^{-2}$ | $9.110 \times 10^{-3}$ |
| Green et al. (2015) | 80 s | 2,000 | Bayestar15 | 0.000000 |
| Green et al. (2018) | 160 s | 2,000 | Bayestar17 | 0.000000 |
| Green et al. (2019) | 130 s | 2,000 | Bayestar19 | 0.000000 |
| Lenz et al. (2017) | 40 s | 2,000 | $E(B-V)$ | $2.834 \times 10^{-6}$ |
| Soto et al. (2019) | 12 s | 555 | $A_K$ | 0.000000 |
| Chen et al. (2019) | 5 m | 1,000 | $A_G$ | $2.099 \times 10^{-4}$ |